\documentclass[aps,pre,reprint,superscriptaddress]{revtex4-1}

\usepackage{amsmath}
\usepackage{txfonts}

\usepackage[T1]{fontenc}
\usepackage{textcomp}
\usepackage[utf8]{inputenc}

\usepackage[pdftex]{graphicx}
\graphicspath{{PercolationThresholdPowerLawPictures/}}

\usepackage{enumerate}

\usepackage[table,dvipsnames]{xcolor}
\definecolor{grey}{gray}{0.70}
\definecolor{grey90}{gray}{0.90}

\usepackage{bm}
\bmdefine{\aVector}{a}
\bmdefine{\AVector}{A}
\bmdefine{\bVector}{b}
\bmdefine{\BVector}{B}
\bmdefine{\cVector}{c}
\bmdefine{\CVector}{C}
\bmdefine{\eVector}{e}
\bmdefine{\EVector}{E}
\bmdefine{\fVector}{f}
\bmdefine{\FVector}{F}
\bmdefine{\gVector}{g}
\bmdefine{\pVector}{p}
\bmdefine{\PVector}{P}
\bmdefine{\qVector}{q}
\bmdefine{\QVector}{Q}
\bmdefine{\rVector}{r}
\bmdefine{\RVector}{R}
\bmdefine{\sVector}{s}
\bmdefine{\uVector}{u}
\bmdefine{\vVector}{v}
\bmdefine{\VVector}{V}
\bmdefine{\muVector}{\mu}
\bmdefine{\OmegaVector}{\Omega}

\begin{document}


\title{Molecular-shape- and size-independent power-law dependence of percolation thresholds on radius of gyration in ideal molecular systems}


\author{Yuki Norizoe}
\email[E-mail: ]{norizoe@cmpt.phys.tohoku.ac.jp}

\author{Toshihiro Kawakatsu}
\affiliation{Department of Physics, Graduate School of Science, Tohoku University - 980-8578 Sendai, Japan}

\author{Hiroshi Morita}
\affiliation{National Institute of Advanced Industrial Science and Technology (AIST) - Central 2-1, 1-1-1 Umezono, Tsukuba, Ibaraki 305-8568, Japan}


\date{January 8, 2021}

\begin{abstract}
Three-dimensional single-component ideal gas systems composed of model homogeneous rigid molecules in various molecular shapes and sizes are simulated by a molecular Monte Carlo simulation technique. We reveal that percolation thresholds of such single-component systems result in, when the molecular volume is fixed, power-law decreasing functions of the radius of gyration (gyradius) of the molecules. The systems with the same parameter set of the molecular volume and radius of gyration, but in different molecular shapes, show the identical value of the percolation threshold. Moreover, we also reveal that a dimensionless scale-free parameter, which is the ratio between the radius of gyration and real cube root of the molecular volume, uniquely determines the percolation threshold.
\end{abstract}


\maketitle

\section{Introduction}
\label{sec:Introduction}
Percolation theory, an important research field of statistical physics, has been studied extensively over the last several decades. Percolation phenomena are observed in and related to various physical, chemical, and social systems and phenomena, such as charge transport, mercury porosimetry, gelation, polymer films, string-like colloidal assembly, communication networks, and epidemics of infectious diseases~\cite{Zallen:PhysicsOfAmorphousSolids,Stauffer1985,Sahimi:ApplicationsOfPercolationTheory,Odagaki:IntroductionToPercolationPhysics,Baranovskii:2019,Norizoe:2019,Norizoe:2014JCP,Norizoe:2012JCP,Norizoe:2005}. The percolation theory also performs an important role in materials science and industry, such as material design and crude oil production. Percolation transition and concomitant critical phenomena, \textit{i.e.} the universal phase behaviour, particularly attract broad attention in the scientific fields~\cite{Stanley:1977,Zallen:PhysicsOfAmorphousSolids,Stauffer1985,Sahimi:ApplicationsOfPercolationTheory,Odagaki:IntroductionToPercolationPhysics,Norizoe:2019,Norizoe:2014JCP,Norizoe:2012JCP,Norizoe:2005}. On the other hand, percolation thresholds themselves also play a key role: percolation thresholds of molecular systems are directly linked to material properties and physical phenomena such as thermal conductivity, electrical resistivity, and magnetism. However, the percolation thresholds in the molecular systems composed of various molecular species, to the best of our knowledge, have not extensively been studied yet.

In the present work, we cast light on this problem. Using molecular Monte Carlo simulation in the 3 dimensional (3-D) continuous space, we study the percolation thresholds of, as the most basic and simplest model molecular system, single-component ideal gases consisting of rigid molecules with the homogeneous intramolecular structure in a variety of molecular shapes and sizes. We reveal that, when the volume of one molecule, denoted by $V_\text{mol}$, is fixed, the percolation thresholds of such single-component model systems result in power-law decreasing functions of $r_\text{g}$, where $r_\text{g}$ denotes the radius of gyration of the molecules. The single-component systems at the identical parameter set of $V_\text{mol}$ and $r_\text{g}$, but with distinct molecular shapes, show the same value of the percolation threshold. Furthermore, we also reveal that a dimensionless scale-free parameter, $r_\text{g} / L_\text{mol}$, uniquely determines the percolation threshold, where $L_\text{mol} = \left( V_\text{mol} \right)^{1/3}$. In brief, here in the present work, we find and study another universal phase behaviour, which is distinct from the above well-known transition and critical phenomena, of percolation systems. Our findings facilitate the prediction of the percolation thresholds, which is closely related to, for example, accident prevention against dust explosion, forest fire, \textit{etc.} This prediction also advances material design. The present work enhances physical understanding of the percolation systems and phenomena.

Percolation thresholds of single-component systems composed of solid (homogeneous and continuous) objects were also studied in early works~\cite{Zallen:PhysicsOfAmorphousSolids,Stauffer1985,Sahimi:ApplicationsOfPercolationTheory,Odagaki:IntroductionToPercolationPhysics,Scher:1970,Webman:1976,Smith:1979,Pike:1974,Alon:1990,Garboczi:1995,Baker:2002}. For example, some early researchers intensively studied systems composed of isotropic objects, \textit{i.e.} circles in 2 dimensions (2-D) and spheres in 3-D. They revealed universal percolation thresholds valid in such systems. In contrast, universal behaviour of percolation thresholds in systems composed of anisotropic objects in various shapes and sizes was reserved for future works. Here we shed light on this problem and reveal a universal relation between percolation thresholds in single-component systems consisting of rigid molecules in different shapes and sizes.

\section{Molecular design}
\label{sec:MolecularDesign}
\begin{figure}[!tb]
	\centering
		\includegraphics[clip,width=8cm]{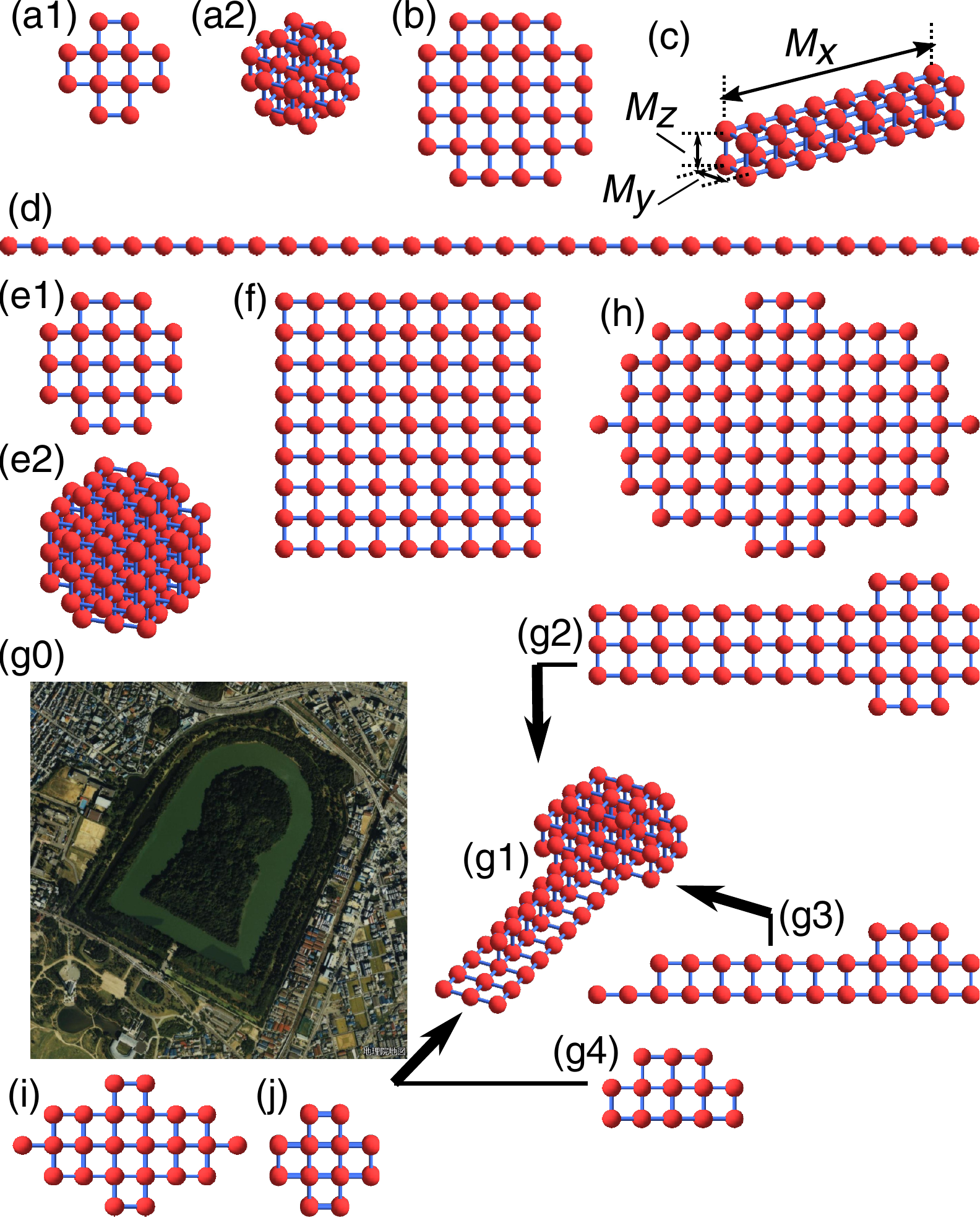}
	\caption{Snapshots of some molecular species. Spherical beads in each panel represent the position of each segment. (a) - (d), $M = 32$. (a): 3-D sphere ((a1) front and (a2) oblique views, respectively), (b): 2-D circle, (c): 3-D cuboid, and (d): 1-D linear molecules. (e) - (h), $M = 81$. (e): 3-D sphere ((e1) front and (e2) oblique views, respectively), (f): 2-D square, (g): 3-D zenp\={o}-k\={o}en-fun, and (h): 2-D ellipse molecules.  (g1): oblique, (g2): top, (g3): side, and (g4): front views of the zenp\={o}-k\={o}en-fun, respectively. Zenp\={o}-k\={o}en-fun means a keyhole-shaped mounded tomb built in ancient Japan. An aerial photo of Daisen Kofun (\copyright Geospatial Information Authority of Japan), which is several hundred meters long and wide and known as a typical zenp\={o}-k\={o}en-fun, is shown in panel (g0) for reference. (i) - (j), $M = 216$. (i): 3-D elliptic cylinder and (j): 3-D cylinder molecules (only top views).}
	\label{fig:SnapshotsIdealRigidMoleculesN216N81N32}
\end{figure}

We design various rigid molecules in different shapes and sizes, \textit{i.e.} molecular species, and simulate single-component systems of these designed molecular species. Here this molecular design is discussed. Rigid molecules, whose molecular conformation (shape) and intramolecular mass distribution are fixed, consist of single-component segments (particles), and can freely move and rotate in the continuous space in the present simulation system. Each molecular species, however, is designed as a single simply connected network on 3-D simple cubic lattice with the grid spacing $\varDelta L$. This network corresponds to a molecular backbone. Single-component segments (particles) are placed on all the lattice points in the network, and each pair of adjacent segments is bound with a rigid bond. These intramolecular network (molecular backbone), segments, and rigid bonds result in one molecular architecture, \textit{i.e.} a molecular species. Snapshots of some examples of the resulting molecular species are given in Fig.~\ref{fig:SnapshotsIdealRigidMoleculesN216N81N32}. The number of segments or lattice points in the molecule, denoted by $M$, corresponds to the molecular weight or degree of polymerization in polymer science. The 1-dimensional (1-D) size of each segment equals $\varDelta L$, so that the molecular volume is defined as,
\begin{equation}
	\label{eq:MolecularVolume}
	V_\text{mol} = M \left( \varDelta L \right)^3.
\end{equation}
When $M$ is fixed, $V_\text{mol}$ is also fixed and independent of the molecular shape, \textit{i.e.} the intramolecular network structure based on the lattice.

For example, when a rectangular parallelepiped region with the size $( M_x, M_y, M_z )$ on the cubic lattice is chosen for the molecular backbone, a rigid rectangular parallelepiped molecule, tentatively referred to as a cuboid molecule, with this size is built. $M = M_x M_y M_z$ for this molecule. As an example, a snapshot of a cuboid molecule with $( M_x = 8, M_y = 2, M_z = 2 )$ is given in Fig.~\ref{fig:SnapshotsIdealRigidMoleculesN216N81N32}(c).

The other molecules simulated in the present work are carved from the cuboid molecules. In other words, all the lattice points inside the molecules are filled with the segments. No hollow molecules are built. All the molecules are homogeneously and fully filled, and discretized according to the cubic lattice. For example, removing all the edges and vertices from cubic molecules with $( M_x = M_y = M_z = 4 )$ and $( M_x = M_y = M_z = 5 )$, one obtains spherical molecules at $M = 32$ and 81 displayed in Figs.~\ref{fig:SnapshotsIdealRigidMoleculesN216N81N32}(a) and (e), respectively. A molecule containing a 3-D intramolecular network is referred to as a 3-D molecule.

The cuboid molecules at $M_z = 1$ contains planar (2-dimensional, 2-D) rectangular molecular configurations. A square molecule with $( M_x = M_y = 9 )$ is shown in Fig.~\ref{fig:SnapshotsIdealRigidMoleculesN216N81N32}(f) as an example of the 2-D molecules. The other molecules, as well as the rectangular ones, which can be carved from the rectangular molecules are tentatively referred to as 2-D molecules. For example, removing all the vertices from a square molecule at $( M_x = M_y = 6 )$, one obtains a circular molecule at $M = 32$ in Fig.~\ref{fig:SnapshotsIdealRigidMoleculesN216N81N32}(b).

Cuboid molecules with $( M_x = M, M_y = M_z = 1 )$, \textit{i.e.} straight molecules, are tentatively referred to as linear (1-D) molecules. A linear molecule at $M = 32$ is displayed in Fig.~\ref{fig:SnapshotsIdealRigidMoleculesN216N81N32}(d) as an example.

In addition to the snapshots of some molecular species in Fig.~\ref{fig:SnapshotsIdealRigidMoleculesN216N81N32}, molecular architecture of all the molecular species simulated in the present work are listed in the 1st and 2nd columns of Table~\ref{table:ListOfMolecularSpecies}. The molecules are designed at $M = 32, 81, \text{ and } 216 \, (= \!\! 6^3)$. $R_\text{e} = 6 \varDelta L$ is defined as the unit length in the simulation system.
\begin{table}[!htb]
	\caption[]{List of the properties of all the molecular species simulated in the present work. $( M_x, M_y, M_z )$ and $( M_x, M_y )$ in the first column denote 3-D cuboid and 2-D rectangular molecules with these sizes, respectively. The figure number is also given in the first column if snapshots of the molecule are available in Fig.~\ref{fig:SnapshotsIdealRigidMoleculesN216N81N32}. The molecules are grouped according to $M$, and ordered by $r_\text{g}'$ in each group. $M = 32, 81$, and 216. \colorbox{lightgray}{\phantom{a}} and \colorbox{grey90}{\phantom{a}} represent 2 and 1-D molecular species, respectively. ``cyl.'' denotes an abbreviation for a cylinder, and ``ZKF'' represents zenp\={o}-k\={o}en-fun.}
	\label{table:ListOfMolecularSpecies}
	\centering
	\begin{tabular}{cclcl} \hline
		shape (and Fig.) & $r_\text{g}'$ & \multicolumn{1}{c}{$\rho_{\text{p}*}'$} & $r_\text{g} / L_\text{mol}$ & \multicolumn{1}{c}{$\rho_{\text{p}*} V_\text{mol}$}  \\ \hline \hline
		\multicolumn{1}{l}{$M = 32$} \\
		sphere, Fig.~\ref{fig:SnapshotsIdealRigidMoleculesN216N81N32}(a) & 0.250 & 1.01 & 0.472 & 0.150  \\  
		\rowcolor{lightgray} circle, (b) & 0.373 & 0.61 & 0.704 & 0.090  \\  
		$( 8, 2, 2 )$, (c) & 0.400 & 0.65 & 0.755 & 0.096  \\  
		\rowcolor{lightgray} $( 8, 4 )$ & 0.425 & 0.52 & 0.803 & 0.077  \\  
		\rowcolor{lightgray} $( 16, 2 )$ & 0.773 & 0.26 & 1.461 & 0.039  \\  
		\rowcolor{grey90} linear, (d) & 1.539 & 0.085 & 2.908 & 0.013  \\  
		\hline
		
		\multicolumn{1}{l}{$M = 81$} \\
		sphere, (e) & 0.342 & 0.47 & 0.475 & 0.176  \\  
		$( 9, 3, 3 )$ & 0.471 & 0.33 & 0.654 & 0.124  \\  
		\rowcolor{lightgray} $( 9, 9 )$(square),(f) & 0.609 & 0.18 & 0.844 & 0.068  \\  
		ZKF, (g) & 0.614 & 0.22 & 0.852 & 0.083  \\  
		\rowcolor{lightgray} ellipse, (h) & 0.618 & 0.18 & 0.857 & 0.068  \\  
		\rowcolor{lightgray} $( 27, 3 )$ & 1.305 & 0.074 & 1.810 & 0.028  \\  
		\rowcolor{grey90} linear & 3.897 & 0.0089 & 5.404 & 0.0033  \\  
		\hline
		
		\multicolumn{1}{l}{$M = 216$} \\
		$( 6, 6, 6 )$ (cube) & 0.493 & 0.175 & 0.493 & 0.175  \\  
		$( 9, 6, 4 )$ & 0.549 & 0.165 & 0.549 & 0.165  \\  
		elliptic cyl., (i) & 0.555 & 0.150 & 0.555 & 0.150  \\  
		$( 9, 8, 3 )$ & 0.591 & 0.140 & 0.591 & 0.140  \\  
		$( 12, 6, 3 )$ & 0.656 & 0.125 & 0.656 & 0.125  \\  
		$( 12, 9, 2 )$ & 0.723 & 0.100 & 0.723 & 0.100  \\  
		cyl., (j) & 0.894 & 0.090 & 0.894 & 0.090  \\  
		$( 18, 4, 3 )$ & 0.895 & 0.095 & 0.895 & 0.095  \\  
		$( 18, 6, 2 )$ & 0.914 & 0.080 & 0.914 & 0.080  \\  
		\rowcolor{lightgray} $( 18, 12 )$ & 1.039 & 0.045 & 1.039 & 0.045  \\  
		$( 24, 3, 3 )$ & 1.170 & 0.065 & 1.170 & 0.065  \\  
		$( 27, 4, 2 )$ & 1.314 & 0.055 & 1.314 & 0.055  \\  
		\rowcolor{lightgray} $( 27, 8 )$ & 1.353 & 0.0375 & 1.353 & 0.0375  \\  
		$( 36, 3, 2 )$ & 1.739 & 0.0350 & 1.739 & 0.0350  \\  
		\rowcolor{lightgray} $( 36, 6 )$ & 1.755 & 0.0275 & 1.755 & 0.0275  \\  
		$( 54, 2, 2 )$ & 2.600 & 0.0125 & 2.600 & 0.0125  \\  
		\rowcolor{lightgray} $( 54, 4 )$ & 2.604 & 0.0125 & 2.604 & 0.0125  \\  
		\rowcolor{lightgray} $( 72, 3 )$ & 3.466 & 0.0085 & 3.466 & 0.0085  \\  
		\rowcolor{lightgray} $( 108, 2 )$ & 5.197 & 0.0040 & 5.197 & 0.0040  \\  
		\hline \hline
		shape (and Fig.) & $r_\text{g}'$ & \multicolumn{1}{c}{$\rho_{\text{p}*}'$} & $r_\text{g} / L_\text{mol}$ & \multicolumn{1}{c}{$\rho_{\text{p}*} V_\text{mol}$}  \\ \hline
	\end{tabular}
\end{table}

\section{Radius of gyration (gyradius)}
\label{sec:RadiusOfGyrationGyradius}
Radius of gyration $r_\text{g}$ is defined as the root mean square distance between the centre of mass of the molecule, denoted by $\rVector_\text{centre}$, and each segment in the molecule~\cite{Doi:IntroductionToPolymerPhysics,Kawakatsu:StatisticalPhysicsOfPolymersAnIntroduction}:
\begin{equation}
	\label{eq:Gyradius}
	r_\text{g} := \left\lbrace \frac{1}{M} \sum_{i=1}^M \left( \rVector_i - \rVector_\text{centre} \right)^2 \right\rbrace ^{1/2}, \quad \rVector_\text{centre} := \frac{1}{M} \sum_{i=1}^M \rVector_i,
\end{equation}
where the summation runs over all the segments in each molecule and $\rVector_i$ denotes the position of the $i$-th segment. $r_\text{g}$, a unique constant of each molecular species, is a measure of the spatial range of the molecule around the centre of mass. Elongation of a molecule with the fixed molecular volume increases $r_\text{g}$ of this molecule. This illustrates that $r_\text{g}$ contains the information on the molecular shape as well as the 1-dimensional size. We utilize this $r_\text{g}$ as a measure of the molecular shapes and sizes. In other words, measuring  $r_\text{g}$ of fully and homogeneously filled molecules, in the present work, we quantify the molecular shapes and sizes. $r_\text{g}$ is reduced to dimensionless quantity, $r_\text{g}' = r_\text{g} / R_\text{e}$, which is given in the 2nd column of Table~\ref{table:ListOfMolecularSpecies}.

\section{Simulation methods}
\label{sec:SimulationMethods}
\begin{figure}[!tb]
	\centering
		\includegraphics[clip,width=8cm]{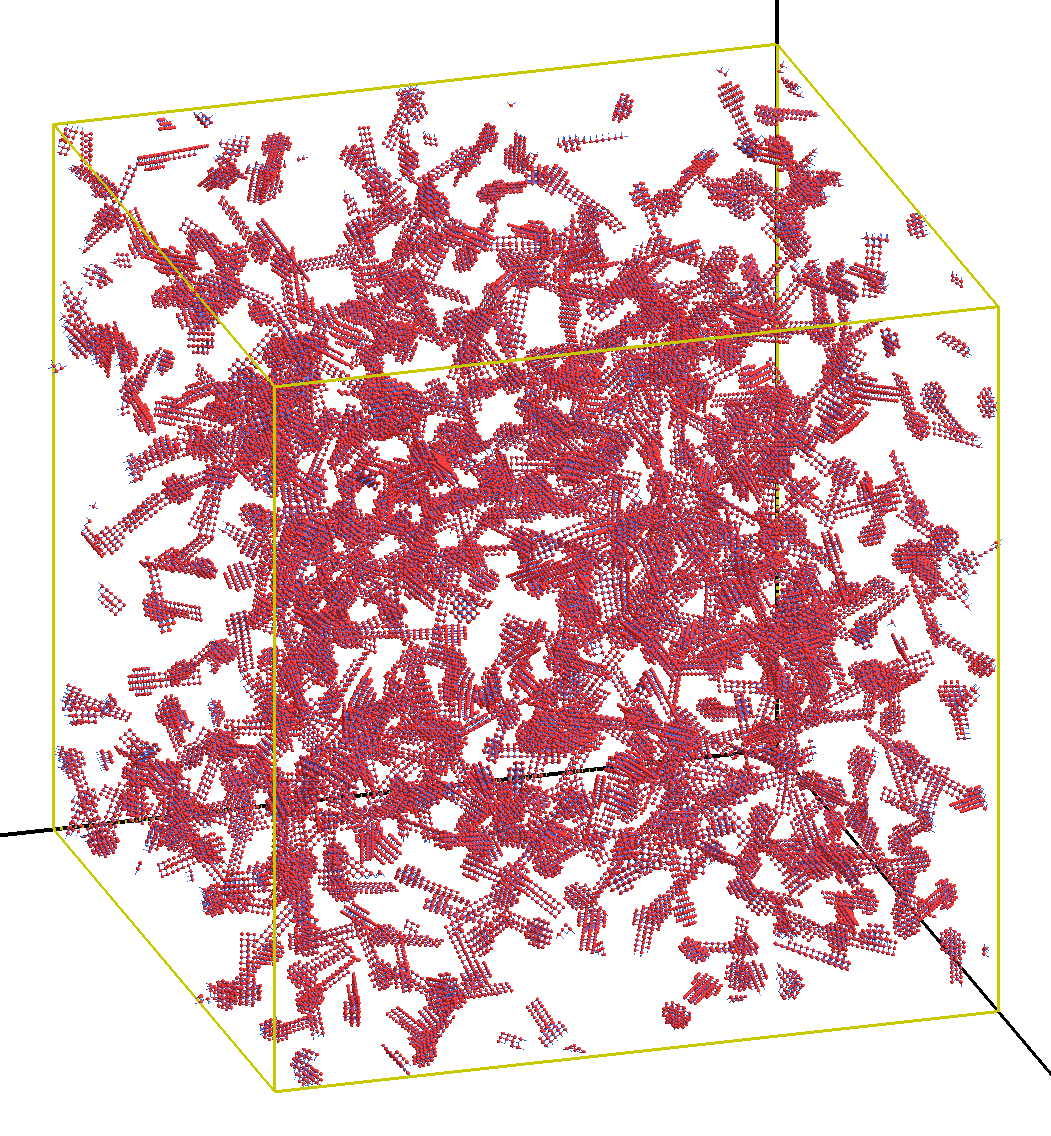}
	\caption{An example of a constructed simulation system. A snapshot of the system composed of zenp\={o}-k\={o}en-fun molecules. $\rho_\text{p}' = 0.1, M = 81$.}
	
	\label{fig:IdealRigidZenpoukouenfunN81ReN32Nx5Ny13Nz3L200R01_000010000MCS}
\end{figure}

The molecules are distributed over a cubic system box of edge length $L_\text{system} = 20 R_\text{e}$ in each 3-D single-component system. This system box, to which a periodic boundary condition is applied, is laid in a spatial region of $0 \le x, y, z < L_\text{system}$. $\rho_\text{p}$ denotes the average volumetric number density of the molecules in the system and $\rho_\text{p}' = \rho_\text{p} R_\text{e}^3$ is the dimensionless average density. For example, the number of molecules in the system box equals $\rho_\text{p} L_\text{system}^3 = 8000$ at $\rho_\text{p}' = 1$. We simulate each single-component system at various values of $\rho_\text{p}'$ and collect 10 independent samples of particle configurations at each value of $\rho_\text{p}'$ in each system. Both the position and direction of each penetrable (permeable) molecule are randomly determined in each sample of the ideal single-component systems. $\rVector_\text{centre}$ of each molecule is uniformly distributed over the system box. Uniformly distributed points on the surface of a unit sphere are used for the direction of the molecules. An example of a snapshot of a constructed simulation system is displayed in Fig.~\ref{fig:IdealRigidZenpoukouenfunN81ReN32Nx5Ny13Nz3L200R01_000010000MCS}.

Occurrence probability of large clusters (percolation clusters) that span the system is measured at each value of $\rho_\text{p}'$ in each system. The transition density (percolation threshold), denoted by $\rho_{\text{p}*}$, in each system is quickly evident because, when $\rho_\text{p}'$ is raised from extremely low values, this occurrence probability abruptly rises from 0 to 1 at $\rho_\text{p}' = \rho_{\text{p}*}'$, where $\rho_{\text{p}*}' = \rho_{\text{p}*} R_\text{e}^3$.

We have verified that the physical properties of the systems are not significantly changed even when larger systems at $L_\text{system} = 40 R_\text{e}$ are simulated, and that $\rho_{\text{p}*}'$ is determined within a relative error of $\approx \! \text{(several \%)}$ in the present simulation system and conditions.

\section{Definition of percolation cluster}
\label{sec:DefinitionOfPercolationCluster}
The percolation cluster is defined \textit{via} a collocation lattice, which is similar to the one defined in our recent works~\cite{Norizoe:2014JCP,Norizoe:2019}. First, the system box is partitioned into a cubic lattice of grid spacing $\varDelta L$, which equals the lattice constant for the intramolecular grids. The index of the resulting small cell is denoted by $( i_x, i_y, i_z )$ for $0 \le i_\alpha < L_\text{system} / \varDelta L$, where $\alpha$ represents the Cartesian coordinates $x, y$, and $z$. We assume that a small cell is occupied when it contains at least one segment. We also assume that a pair of the occupied small cells is linked when this pair satisfies a relation,
\begin{equation}
	\label{eq:LinkDefinition}
	\left| i_\alpha - i'_\alpha \right| \le 1 \quad \text{for} \quad \alpha = x, y, z,
\end{equation}
where $i_\alpha$ and $i'_\alpha$ denote the indices of the pair respectively. A single network of links is referred to as a cluster. A percolation cluster is defined as a cluster bridging both the faces of the system box. The number of small cells equals $( L_\text{system} / \varDelta L )^3 = 120^3 = 1.728 \times 10^6$. This disturbs both a drastic increase in $L_\text{system}$ and finite size scaling.

\section{Simulation results}
\label{sec:SimulationResults}
\begin{figure}[!tb]
	\centering
		\includegraphics[clip]{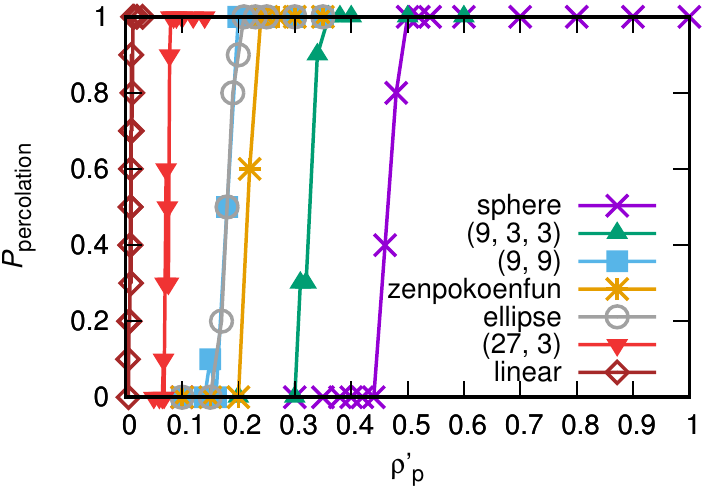}
	\caption{The dependence of $P_\text{percolation}$ on $\rho_\text{p}'$ in each system at $M = 81$, where $P_\text{percolation}$ denotes the occurrence probability of percolation clusters. Although $r_\text{g}'$ of the 3-D zenp\={o}-k\={o}en-fun is larger than $r_\text{g}'$ of the 2-D $( 9, 9 )$ square, the zenp\={o}-k\={o}en-fun exceeds the square in $\rho_{\text{p}*}'$. Artefacts due to the molecular dimensionality and the grid-based measurement of the extramolecular network structure cause this inversion, which is discussed below.}
	\label{fig:IdealRigidN81ReN32L200-Percolation_Publication}
\end{figure}

Simulation results are discussed here. The occurrence probability of percolation clusters is measured at each value of $\rho_\text{p}'$ in each single-component system, as mentioned above. As an example, the dependence of the occurrence probability on $\rho_\text{p}'$ in each system at $M = 81$ is given in Fig.~\ref{fig:IdealRigidN81ReN32L200-Percolation_Publication}. The values of $\rho_{\text{p}*}'$ determined in each system based on these results are listed in the 3rd column of Table~\ref{table:ListOfMolecularSpecies}. $\rho_{\text{p}*}'$ decreases with increasing $r_\text{g}'$ at each value of $M$, which is consistent with an intuitive understanding that elongate molecules can connect both the sides of the system box at low density. $\rho_{\text{p}*}'$ is also reduced when $M$ rises, because large molecules facilitate the percolation.

\begin{figure}[!tb]
	\centering
		\includegraphics[clip,width=8cm]{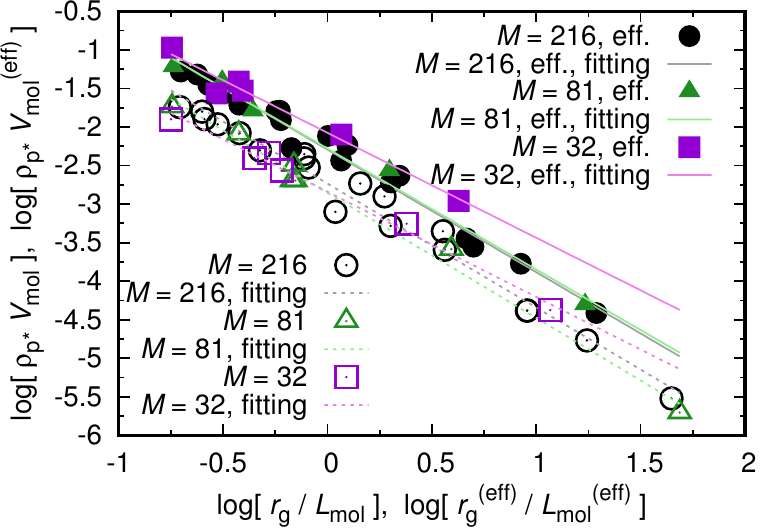}
	\caption{The dependence of $\rho_{\text{p}*}'$ on $r_\text{g}'$ at each value of $M$, represented by open symbols. These quantities are normalized and reduced to dimensionless scale-free quantities $\rho_{\text{p}*} V_\text{mol}$ and $r_\text{g} / L_\text{mol}$, respectively, so that $\log [ r_\text{g} / L_\text{mol} ]$-$\log [ \rho_{\text{p}*} V_\text{mol} ]$ graph is plotted, in which ``$\log$'' denotes the natural logarithm. Broken straight lines represent linear fittings of the graph of these open symbols at each value of $M$, which is $y = -\lambda_\text{N} x + \omega_\text{N}$:
		$( \lambda_\text{N} = 1.4, \omega_\text{N} = -2.8 )$ at  $M = 32$,
		$( \lambda_\text{N} = 1.6, \omega_\text{N} = -2.9 )$ at  $M = 81$, and
		$( \lambda_\text{N} = 1.6, \omega_\text{N} = -2.7 )$ at  $M = 216$.
		On the other hand, filled symbols represent $\log [ r_\text{g}^\text{(eff)} / L_\text{mol}^\text{(eff)} ]$-$\log [ \rho_{\text{p}*} V_\text{mol}^\text{(eff)} ]$ graph for all the molecular species except the ones in Fig.~\ref{fig:SnapshotsIdealRigidMoleculesN216N81N32} (g), (h), and (i). Linear fittings, denoted by solid lines, of the graph of these filled symbols at each value of $M$, which is $y = -\lambda_\text{N} x + \omega_\text{N}$:
		$( \lambda_\text{N} = 1.4, \omega_\text{N} = -2.1 )$ at  $M = 32$,
		$( \lambda_\text{N} = 1.6, \omega_\text{N} = -2.3 )$ at  $M = 81$, and
		$( \lambda_\text{N} = 1.6, \omega_\text{N} = -2.3 )$ at  $M = 216$.
	}
	\label{fig:IdealRigidN216N81N32ReN32L200-PercolationThreshold}
\end{figure}

$\rho_{\text{p}*}'$ as a function of $r_\text{g}'$ is plotted in Fig~\ref{fig:IdealRigidN216N81N32ReN32L200-PercolationThreshold}. $r_\text{g}'$ and $\rho_{\text{p}*}'$ are normalized and reduced to dimensionless scale-free quantities $r_\text{g} / L_\text{mol}$ and $\rho_{\text{p}*} V_\text{mol}$, respectively, in this double logarithmic plot, where ``$\log$'' denotes the natural logarithm. Values of these dimensionless scale-free quantities in each system are given in the 4th and 5th columns in Table~\ref{table:ListOfMolecularSpecies}. These results demonstrate that, when $V_\text{mol}' = V_\text{mol} /  R_\text{e}^3$ is fixed, $\rho_{\text{p}*}'$ results in a power-law decreasing function of $r_\text{g}'$. The single-component systems at the same parameter set of $V_\text{mol}'$ and $r_\text{g}'$, but different in molecular shape, indicate the identical value of $\rho_{\text{p}*}'$.

Moreover, this power-law relation between $r_\text{g} / L_\text{mol}$ and $\rho_{\text{p}*} V_\text{mol}$ is independent of values of $V_\text{mol}'$. In other words, $r_\text{g} / L_\text{mol}$ uniquely determines $\rho_{\text{p}*} V_\text{mol}$. When a value of $r_\text{g} / L_\text{mol}$ is given, $\rho_{\text{p}*} V_\text{mol}$ shows the identical and unique value, independently of both the molecular shape and value of $V_\text{mol}$. Linear fittings of the log-log graph
at each value of $V_\text{mol}'$ indicate,
\begin{equation}
	\label{eq:LinearFittingScaleFreeVolumeFractionAndGyradius}
	y = -\lambda_\text{N} x + \omega_\text{N}, \quad ( \lambda_\text{N} \cong 1.5 \pm 0.1, \omega_\text{N} \cong -2.8 \pm 0.1 ).
\end{equation}
The suffix of $\lambda_\text{N}$ and $\omega_\text{N}$, ``N'', is added to avoid confusion between the present (New) and conventional critical exponents.

These simulation results and physical understanding of $r_\text{g} / L_\text{mol}$ and $\rho_{\text{p}*} V_\text{mol}$ are further discussed below.

\section{Dimensionless scale-free quantities and artefacts due to the lattice}
\label{sec:DimensionlessScaleFreeQuantities}
$r_\text{g}$ characterizes the molecular shape as well as the 1-D size. $L_\text{mol} = \left( V_\text{mol} \right)^{1/3}$ represents another definition of the 1-D molecular size and provides no information on the molecular shape. The ratio between these two parameters, $r_\text{g} / L_\text{mol}$, defines and characterizes the dimensionless scale-free shape of the molecule. $r_\text{g} / L_\text{mol}$ resembles, in this sense, the aspect ratio of rectangles, \textit{i.e.} a dimensionless scale-free parameter. $r_\text{g} / L_\text{mol}$ of any object with similar shapes equals the sole and identical value. As an example, $r_\text{g}$ and $r_\text{g} / L_\text{mol}$ of some solid objects (homogeneous and continuous rigid objects) are listed in Table~\ref{table:ListOfSolidObjects}. On the ohter hand, $\rho_{\text{p}} V_\text{mol}$ is equivalent to another dimensionless scale-free quantity, \textit{i.e.} volume fraction including the duplication of overlapping spatial regions among the molecules (solid objects), or, in other words, volume fraction measured when all the molecules are isolated. Therefore, when the overlaps (links) between solid objects are measured exactly in off-lattice unlike our grid-based measurement, points $( r_\text{g}, \rho_{\text{p}*} )$ determined in single-component systems composed of solid objects with the same shape and different scales of objects are mapped onto the sole and identical point in $( r_\text{g} / L_\text{mol} )$-$( \rho_{\text{p}*} V_\text{mol} )$ plane. Thus each shape is mapped onto one point in $( r_\text{g} / L_\text{mol} )$-$( \rho_{\text{p}*} V_\text{mol} )$ plane in the systems of solid objects. When a large number of different shapes are mapped onto corresponding points, a line (curve) is constructed in $( r_\text{g} / L_\text{mol} )$-$( \rho_{\text{p}*} V_\text{mol} )$ plane. This line, $\rho_{\text{p}*} V_\text{mol}$, becomes a decreasing function of $r_\text{g} / L_\text{mol}$ because elongate solid objects, intuitively, facilitate the percolation. We tentatively refer to this line, which is constructed in the systems of solid objects, as a ``unified'' line or a unified result in the present work. In other words, a function $\rho_{\text{p}*}$ of $r_\text{g}$ determined in such completely continuous single-component systems with various different object shapes at any fixed object scale (molecular volume) is always mapped onto the sole unified line in $( r_\text{g} / L_\text{mol} )$-$( \rho_{\text{p}*} V_\text{mol} )$ plane. Figure~\ref{fig:IdealRigidN216N81N32ReN32L200-PercolationThreshold} shows that the plotted points and linear fittings at $M = 32, 81$, and 216 lie in the vicinity of this unified line, and that the unified result satisfies a relation,
\begin{equation}
	\label{eq:PowerLawVolumeFractionOfScaleFreeGyradius}
	\rho_{\text{p}*} V_\text{mol} \cong \exp [ \omega_\text{N} ] \, \left( r_\text{g} \middle/ L_\text{mol} \right)^{-\lambda_\text{N}},
\end{equation}
which has also been shown in eq.~\eqref{eq:LinearFittingScaleFreeVolumeFractionAndGyradius}.
\begin{table}[!htb]
	\caption[]{List of $r_\text{g}$ and $r_\text{g} / L_\text{mol}$ for some solid objects: a sphere (solid sphere) $R$ in radius, a cuboid with the size $\left( l_x, l_y, l_z \right)$, a cube of edge length $l$, and a circular cylinder $R$ in radius and $h$ in height. $l_\text{diag} = \left( l_x^2 + l_y^2 + l_z^2 \right)^{1/2}$. $r_\text{g,cyl} = \left\lbrace (1/2) R^2 + (1/12) h^2 \right\rbrace ^{1/2}$.}
	\label{table:ListOfSolidObjects}
	\centering
	\begin{tabular}{ccc} \hline
		& $r_\text{g}$ & $r_\text{g} / L_\text{mol}$  \\
		\hline
		sphere & $(3/5)^{1/2} R$ & $(3/5)^{1/2} (4 \pi / 3)^{-1/3} \approx 0.4805$  \\
		cuboid & $12^{-1/2} l_\text{diag}$ & $12^{-1/2} l_\text{diag} \left( l_x l_y l_z \right)^{-1/3}$  \\
		cube & $l / 2$ & $1/2$  \\
		cylinder & $r_\text{g,cyl}$ & $r_\text{g,cyl} \left( \pi R^2 h \right)^{-1/3}$  \\
		\hline
	\end{tabular}
\end{table}

However, the similarity in shapes between our molecules is broken because of the intramolecular discretization. This results in inconsistency of values of $r_\text{g} / L_\text{mol}$ between the molecules at different values of $M$. For example, spherical molecules at $M = 32$ and 81 indicate slightly different values of $r_\text{g} / L_\text{mol}$. As another example,
\begin{align}
	\label{eq:GyradiusOfDicsretizedCube}
	r_\text{g} &= (1/2) ( M_0 - 1 )^{1/2} ( M_0 + 1 )^{1/2} \varDelta L,  \\
	\label{eq:GyradiusOverLmolOfDicsretizedCube}
	r_\text{g} / L_\text{mol} &= \left. (1/2) ( M_0 - 1 )^{1/2} ( M_0 + 1 )^{1/2} \middle/ M_0 \right.,
\end{align}
for cubic molecules at $M_x = M_y = M_z = M_0$ and $M = M_0^3$. These are consistent with the result of solid cubes, listed in Table~\ref{table:ListOfSolidObjects}, for $M_0 \to \infty$, and, by contrast, dependent on $M$ at finite $M_0$. On the other hand, only slight inconsistency of $r_\text{g} / L_\text{mol}$ is found between the discretized molecules which have the similar molecular shapes. The results of discretized/solid spheres, cuboids, and cubes listed in Tables~\ref{table:ListOfMolecularSpecies} and \ref{table:ListOfSolidObjects} also demonstrate insignificance of difference in $r_\text{g} / L_\text{mol}$ between the discretized molecules and corresponding solid objects.

The slight inconsistency between the above unified line and simulation results at $M = 32, 81$, and 216 shown in Fig.~\ref{fig:IdealRigidN216N81N32ReN32L200-PercolationThreshold} is attributed to the grid-based measurement of the extramolecular network structure and $\rho_{\text{p}*}$. $M$ segments of an isolated long $(M \gg 1)$ linear molecule occupy $M$ small cells in the system box when this molecule aligns parallel with one of the $x$-, $y$-, and $z$-axes. By contrast, fewer small cells than $M$ are occupied when this molecule lies non-parallel to all the three axes because two adjacent segments of this molecule can occupy the same small cell. This reduction in the number of occupied small cells equals a decrease in the effective molecular volume, and raises the percolation threshold, $\rho_{\text{p}*}$. This artefact due to the collocation lattice which partitions the system box rises when the total number of nearest and next nearest neighbours in the intramolecular network increases. Therefore, the artefact is raised when the dimensionality of the molecule increases at fixed $M$. These results illustrate that the artefact due to the collocation lattice sensitively depends on the molecular architecture.

The collocation lattice also causes another artefact. As discussed in section~``Molecular design'' and briefly shown in eq.~\eqref{eq:MolecularVolume}, each segment is considered a cube $\varDelta L$ in edge length: a segment is equivalent to a point mass centred in this cubic region owned or occupied by this segment. For example, when a pair of the segments, \textit{i.e.} point masses, is positioned in small cells of $( i_x, i_y, i_z )$ and $( i_x + 1, i_y, i_z )$ respectively, these small cells are linked according to eq.~\eqref{eq:LinkDefinition}, and one assumes that these segments are connected and overlap. However, eq.~\eqref{eq:LinkDefinition} is independent of the positions, inside these small cells, of these segments. These segments overlap at $\varDelta x \approx 2 \varDelta L$ as well as at $\varDelta x \approx 0$ whenever eq.~\eqref{eq:LinkDefinition} is satisfied, where $\varDelta x$ denotes a difference in $x$-coordinates between these segments. Similar results also hold for $y$- and $z$-coordinates.
This illustrates that each segment is considered a cube of edge length $\approx \! 2 \varDelta L$, and increases the effective molecular size by $\approx \! 0.5 \varDelta L$ in the direction normal to the molecular surface. For example, the size of a cuboid molecule rises from $( M_x, M_y, M_z )$, effectively, to $( M_x + 1, M_y + 1, M_z + 1 )$. This increase in the effective molecular size also raises effective values of $( M, V_\text{mol}, L_\text{mol}, r_\text{g}, \rho_{\text{p}*} V_\text{mol} )$ of each molecular species in Table~\ref{table:ListOfMolecularSpecies}, and changes $r_\text{g} / L_\text{mol}$ effectively. By contrast, $\rho_{\text{p}*}'$ is unchanged. For example, these size and physical properties of a spherical molecule at $M = 32$ (Fig.~\ref{fig:SnapshotsIdealRigidMoleculesN216N81N32}(a)) rises to, effectively, those at $M = 81$ (Fig.~\ref{fig:SnapshotsIdealRigidMoleculesN216N81N32}(e)) listed in Table~\ref{table:ListOfMolecularSpecies}, with the fixed $\rho_{\text{p}*}'$ at $M = 32$. Such effective values are calculated for all the molecular species in Table~\ref{table:ListOfMolecularSpecies} except the ones in Fig.~\ref{fig:SnapshotsIdealRigidMoleculesN216N81N32} (g), (h), and (i). $\log [ r_\text{g}^\text{(eff)} / L_\text{mol}^\text{(eff)} ]$-$\log [ \rho_{\text{p}*} V_\text{mol}^\text{(eff)} ]$ graph is plotted in Fig.~\ref{fig:IdealRigidN216N81N32ReN32L200-PercolationThreshold}, where ``${}^\text{(eff)}$'' denotes the calculated effective quantities. The results still satisfy the power-law relation, eq.~\eqref{eq:PowerLawVolumeFractionOfScaleFreeGyradius}, with the identical value of $\lambda_\text{N}$ as eq.~\eqref{eq:LinearFittingScaleFreeVolumeFractionAndGyradius}, and slightly raised $\omega_\text{N}$,
\begin{equation}
	\label{eq:CriticalExponents_lambdaN_omegaN}
	( \lambda_\text{N} \cong 1.5 \pm 0.1, \omega_\text{N} \cong -2.2 \pm 0.1 ).
\end{equation}
This corroborates the validity and universality of the exponents and eq.~\eqref{eq:PowerLawVolumeFractionOfScaleFreeGyradius}.

The points plotted in the graph of Fig.~\ref{fig:IdealRigidN216N81N32ReN32L200-PercolationThreshold} are strewn in the vicinity of the unified line. This is similar to the percolation thresholds in the systems composed of isotropic objects, in which the universal percolation thresholds were found in early works. The percolation thresholds in such systems show, in the vicinity of the universal values, slight and sensitive dependence on details of each system, such as lattice structure \textit{etc.} ~\cite{Zallen:PhysicsOfAmorphousSolids,Stauffer1985,Sahimi:ApplicationsOfPercolationTheory,Odagaki:IntroductionToPercolationPhysics,Scher:1970}.

So far disadvantages of our grid-based simulation system have been elucidated. Next, advantages are clarified below.

\section{Advantages of the present grid-based simulation system}
\label{sec:AdvantagesOfPresentGrid-basedSimulationSystem}
As has been mentioned in the introduction, systems composed of isotropic solid objects were intensively studied in early works, and universal percolation thresholds in such systems were revealed~\cite{Zallen:PhysicsOfAmorphousSolids,Stauffer1985,Sahimi:ApplicationsOfPercolationTheory,Odagaki:IntroductionToPercolationPhysics}. The distance, contacts, and overlaps between the objects (links) are readily evaluated in the systems of isotropic solid objects. By contrast, simple anisotropic shapes as well as complicated ones pose a challenge in the evaluation of links. This could have discouraged extensive and systematic studies on percolation thresholds in single-component systems composed of solid objects in various shapes and sizes. Moreover, parameters such as aspect ratios \textit{etc.} were typically chosen rather than the radius of gyration $r_\text{g}$. The importance and versatility of $r_\text{g}$ were overlooked. Utilizing $r_\text{g}$, which is automatically calculated for any sophisticated molecule, one can quantify any complicated shape and readily compare any quantified shapes in equal conditions. Furthermore, $r_\text{g} / L_\text{mol}$ uniquely determines percolation thresholds according to eq.~\eqref{eq:PowerLawVolumeFractionOfScaleFreeGyradius}. Contrary to $r_\text{g}$, the aspect ratios \textit{etc.} were specifically designed for and applicable to particular shapes. The volume fraction was, by contrast, frequently found, whereas the duplication of overlapping spatial regions between the objects was, in typical early works, eliminated in their definition of the volume fraction when systems composed of penetrable (permeable) objects were studied, unlike the definition of our $\rho_{\text{p}*} V_\text{mol}$ which includes the duplication although our systems consist of penetrable molecules. The average volume fraction of single-component systems composed of randomly distributed penetrable solid objects at the average number density $\rho_{\text{p}}$ in their definition, denoted by $\phi_\text{conventional}$, is,
\begin{equation}
	\label{eq:ConventionalVolumeFraction}
	\phi_\text{conventional} = 1 - \exp( - \rho_{\text{p}} V_\text{mol} ),
\end{equation}
which is independent of object shapes~\cite{Shante:1971,Garboczi:1991}. These differences from the parameters utilized in the present work could have also disturbed the systematic and extensive studies on percolation thresholds in various systems.

Contrary to the early works mentioned above, we have designed the molecular species in a variety of shapes and sizes using the intramolecular lattice. Combination of the present molecular design and grid-based measurement of percolation thresholds significantly facilitates simulation studies on percolation thresholds. We can systematically and readily construct molecular systems composed of various, even sophisticated, molecular species, for example zenp\={o}-k\={o}en-fun, and automatically measure the percolation threshold of each system. We have introduced $r_\text{g} / L_\text{mol}$ and $\rho_{\text{p}*} V_\text{mol}$, and revealed the importance of and relation between these two quantities, eq.~\eqref{eq:PowerLawVolumeFractionOfScaleFreeGyradius}.

Here, results of the single-component system composed of solid spheres are compared between the early and present works. $r_\text{g} / L_\text{mol}$ of solid spheres listed in Table~\ref{table:ListOfSolidObjects} and eqs.~\eqref{eq:PowerLawVolumeFractionOfScaleFreeGyradius} and \eqref{eq:CriticalExponents_lambdaN_omegaN} give $\rho_{\text{p}*} V_\text{mol} \cong 0.33
$. This is translated into $\phi_\text{conventional} = 0.28$ according to eq.~\eqref{eq:ConventionalVolumeFraction}. This value, \textit{i.e.} a result of the present work, of $\phi_\text{conventional}$ is consistent with early results ~\cite{Pike:1974}, $\approx \! \! 0.30$, and corroborates our findings.

\section{Results at large $r_\text{g} / L_\text{mol}$}
\label{sec:ResultsAtLargeRgOverLmol}
We have simulated $\rho_{\text{p}*}$ in regions of finite $r_\text{g} / L_\text{mol}$ and found eq.~\eqref{eq:PowerLawVolumeFractionOfScaleFreeGyradius}. Utilizing a solid circular or non-circular cylinder $h$ in height, here we examine validity of eq.~\eqref{eq:PowerLawVolumeFractionOfScaleFreeGyradius} in regions of extremely large $r_\text{g} / L_\text{mol}$, in other words extremely elongate solid objects and molecules. The area of a base, or a cross section, of this cylinder is denoted by $A$, and $V_\text{mol} = A h$. $L_\text{mol} = (A h)^{1/3}$. When this cylinder is placed parallel to $z$-axis and fixed at $\rVector_\text{centre} = 0$, $r_\text{g}^2 = (1/12) h^2 + r_\text{g,base}^2$,
where $r_\text{g,base}$ denotes the radius of gyration in a base (or a cross section) of the cylinder, $r_\text{g,base} = \left\lbrace  (1/A) \iint_S \left[ x^2 + y^2 \right] \, dx \, dy \right\rbrace ^{1/2}$.
$r_\text{g} / L_\text{mol}$ of this cylinder, eq.~\eqref{eq:PowerLawVolumeFractionOfScaleFreeGyradius}, and a relation $\lambda_\text{N} \cong 1.5$ result in,
\begin{equation*}
	\label{eq:RhopVmolOfSolidNoncircularCylinderObject}
	\rho_{\text{p}*} V_\text{mol} \cong \exp [ \omega_\text{N} ] \, \left\lbrace (1/12) + \left( 1 \middle/ h^2 \right)  r_\text{g,base}^2 \right\rbrace ^{-3/4} \frac{ A^{1/2} }{h}.
\end{equation*}
This provides a relation for $r_\text{g,base} \ll h$, \textit{i.e.} for extremely elongate cylinders,
\begin{equation}
	\label{eq:RhopVmolOfSolidNoncircularCylinderObjectExtremelyLong}
	\rho_{\text{p}*} V_\text{mol} \cong \exp [ \omega_\text{N} ] \cdot 12^{3/4} ( A^{1/2} / h ) .
\end{equation}
This is the result in regions of extremely large $r_\text{g} / L_\text{mol}$. For example, when $r_\text{g} / L_\text{mol}$ of a solid circular cylinder, or a solid capsule (spherocylinder), $R$ in radius and $h$ in cylindrical height is given to eq.~\eqref{eq:PowerLawVolumeFractionOfScaleFreeGyradius}, $\rho_{\text{p}*} V_\text{mol} \cong \exp [ \omega_\text{N} ] \cdot 12^{3/4} \pi^{1/2} ( R / h )$ for $R \ll h$. This relation is consistent with a known result of capsules~\cite{Schilling:2015,Bug:1985_PRL_1}, $\rho_{\text{p}*} V_\text{mol} \propto R / h$ for $R \ll h$, as well as eq.~\eqref{eq:RhopVmolOfSolidNoncircularCylinderObjectExtremelyLong}. This known result of capsules was also analytically predicted based on connectedness percolation theory~\cite{Kyrylyuk:2008}. These results demonstrate that eq.~\eqref{eq:PowerLawVolumeFractionOfScaleFreeGyradius} is also valid in regions of large $r_\text{g} / L_\text{mol}$.

\section{Conclusions}
\label{sec:Conclusions}
In conclusion, introducing $r_\text{g} / L_\text{mol}$ and $\rho_{\text{p}*} V_\text{mol}$, we have studied percolation thresholds in single-component ideal gas systems composed of rigid molecules in various shapes and sizes, and revealed eq.~\eqref{eq:PowerLawVolumeFractionOfScaleFreeGyradius}. This relation between $r_\text{g} / L_\text{mol}$ and $\rho_{\text{p}*} V_\text{mol}$ is universally satisfied in these systems. Furthermore, our results are, while our simulation system consists of penetrable objects, consistent with results of analytical studies on systems composed of deformed and polygonal rods \textit{etc.} based on connectedness percolation theory~\cite{Drwenski:2017,Drwenski:2018}.

We recently simulated percolation phenomena in systems consisting of interacting flexible molecules using a solvent-free coarse-grained model~\cite{Norizoe:2019,Norizoe:2014JCP}. The present ideal system is constructed based on these recently published more realistic model systems, so that we can expand the present work into the systems consisting of interacting molecules using the solvent-free model, and directly compare the results between the ideal and realistic systems. However, this is beyond the scope of the present work and will be discussed in our forthcoming article.



\begin{acknowledgments}
	This work is based on results obtained from a project, JPNP16010, commissioned by the New Energy and Industrial Technology Development Organization (NEDO).
\end{acknowledgments}

%

\end{document}